\documentclass[aps,prl,reprint,twocolumns,notitlepage]{revtex4-1}
\usepackage{amssymb,bm}
\usepackage{amsmath,eucal,ulem}
\usepackage[usenames,dvipsnames]{color}
\usepackage{graphicx}
\usepackage{units}

\newcommand*{\VEC}[1]{\boldsymbol{#1}}
\newcommand*{\TENSOR}[1]{\mathsf{#1}}

\newcommand*{\kB}{k_{\mathrm{B}}}

\DeclareMathOperator{\erfc}{erfc}

\begin{document}

\title{Driven anisotropic diffusion at boundaries: noise rectification and particle sorting}
\author{Stefano Bo}
\author{Ralf Eichhorn}
\affiliation{Nordita, Royal Institute of Technology and Stockholm University,
Roslagstullsbacken 23, SE-106 91 Stockholm, Sweden}
\begin{abstract}
We study the diffusive dynamics of a Brownian particle
in proximity of a flat surface
under non-equilibrium conditions, which are created by an
anisotropic thermal environment with different temperatures being active
along distinct spatial directions.
By presenting the exact time-dependent solution of the Fokker-Planck equation for this problem,
we demonstrate that the
interplay between anisotropic diffusion and hard-core interaction with the plain wall
rectifies the thermal fluctuations and induces directed particle transport parallel to the surface,
without any deterministic forces being applied in that direction.
Based on current micromanipulation technologies, we suggest a concrete experimental set-up to observe
this novel noise-induced transport mechanism.
We furthermore show that it is sensitive to
particle characteristics, such that this set-up can be used for sorting particles of different size.
\end{abstract}
\date{\today}
\maketitle

\paragraph*{Introduction.}
The ability to manipulate, monitor and fabricate microscopic 
systems has witnessed a dramatic increase in recent years,
opening the way towards accurate analysis of physical processes
on scales where thermal fluctuations are lead actors. 
A key finding of these developments is that such fluctuations are not
necessarily a detrimental nuisance,
but rather may provide novel, noise-induced mechanisms for
controlling motion and performing specific tasks at the microscale.
Well-known examples are
ratchets rectifying fluctuations by means of (spatial or time-reversal) asymmetries
\cite{buttiker87,landauer88,reimann02,reimann02a,hanggi09},
Brownian engines extracting work from fluctuations
\cite{blickle12,martinez16_carnot,martinez16},
and microscopic Maxwell demons exploiting fluctuations to
convert information into energy
\cite{parrondo15,lutz15}.

In the present Letter we suggest a new strategy for rectifying thermal noise
into directed movement of a
microscopic particle, which requires only three simple ingredients:
an anisotropic thermal environment, a plain, unstructured surface (or ``hard wall''),
and a static force oriented towards the surface so that particle
motion is constrained to its close vicinity.
The anisotropic thermal environment induces anisotropic diffusion,
even for a spherical particle.
As we will show below, the interplay between this diffusive behavior
and the constraining boundary couples the drift components
of the particle motion
perpendicular and parallel to the surface.
As a consequence, the particle on average moves along the surface,
even though there is no deterministic force component applied in that direction.
The drift velocity of that directed
movement is essentially controlled by the anisotropy of
the bath.
Unlike common noise-rectifying ratchets, this mechanism
does therefore not require any
state-dependent diffusion \cite{buttiker87,landauer88},
spatially asymmetric periodic force field or   ratchet-like
topographical structure, or any time-asymmetric driving forces
\cite{reimann02,reimann02a,hanggi09}.

From an experimental viewpoint, an 
anisotropic thermal environment may be realized, for instance,
by two different
temperature ``baths'' acting along different directions, or
by a superposition of the usual 
fluid bath of the Brownian particle with a second, ``hotter''
source of (almost) white-noise fluctuations applied along a
specific direction \cite{filliger07}.
In a number of recent experiments, the latter alternative has been implemented
by means of   noisy electrostatic fields
\cite{gomez10,martinez13,mestres14,berut14,dieterich15,martinez15,martinez16_carnot,berut16,martinez16,dinis16,soni17}.
This technique has been shown to provide effective heatings to extremely high temperatures,
and it has been applied to implement a microscale heat engine \cite{martinez16_carnot,martinez16,dinis16}.
It thus appears to be an ideal candidate for creating the 
anisotropic thermal environment needed for establishing non-equilibrium
conditions in our system.
We analyze such an experimental set-up and demonstrate
how the noise rectification can be exploited
to transport differently-sized colloidal beads
in opposite directions along the surface.
This setup thus
provides an elegant method for efficient particle sorting
  without feedback control,
which can be implemented at low cost with state-of-the-art microfluidics
technology \cite{martinez16_carnot}.

Stochastic particle motion on the microscale is commonly modeled in terms
of an overdamped Langevin equation (neglecting inertia effects) \cite{mazo02,snook07}.
The presence of a reflecting wall 
requires additional care to correctly account for the hard-core interactions with these boundaries,
as they cannot directly be introduced into the equation of motion as well-defined interaction forces
\cite{behringer11,behringer12}.
However, in the equivalent representation of driven diffusion in terms of the Fokker-Planck equation
\cite{risken84},
governing the time evolution of the probability density of particle positions,
hard walls are easily implemented as reflecting boundary conditions.
In one dimension, such Fokker-Planck equation with
a reflecting boundary (i.e.~a ``hard wall'' constraining
particle diffusion to the one-dimensional half-space)
has been solved analytically
by Smoluchowski \cite{smoluchowski17,chandrasekhar43},
even in the presence of a constant external force.
He obtained an expression for the time-dependent probability density
in terms of exponential and error functions (see Eq.~\eqref{eq:px3} below).

For isotropic thermal baths,
Smoluchowski's result is immediately generalized to particle diffusion close to
a reflecting surface in three (and also higher) dimensions, because the
different spatial directions are uncorrelated, so that the
three-dimensional Fokker-Planck equation decouples into a set of
three one-dimensional equations.
However,
for anisotropic diffusion with principal directions
not being aligned with
the surface,
correlations between the spatial components perpendicular and
parallel to the surface prohibit such a simple decomposition.
To the best of our knowledge, the analytical solution for this
situation---anisotropic particle diffusion driven by a constant force
in the proximity of a hard surface---is not known in the literature.
We here present the exact time-dependent analytical solution
of the corresponding Fokker-Planck equation,
and reveal that it is intimately connected to the original Smoluchowski
solution for the one-dimensional case.
The above mentioned theoretical predictions of noise-induced
systematic particle motion along the surface
are derived from the properties of this solution.

\paragraph*{Model.}
We model the particle's diffusive motion
in three dimensions
using the Fokker-Planck
equation \cite{risken84}
for the probability density $p(t;\VEC{x})$
of finding the particle at a given position
$\VEC{x}=(x_1,x_2,x_3)$
at time $t$,
\begin{equation}
\label{eq:FP}
\frac{\partial p}{\partial  t}=
-\frac{\partial}{\partial x_i} \left( v_i - \frac{\partial}{\partial x_j} D_{ij} \right) p \, ,
\end{equation}
where summation over repeated indices is understood.
The constant drift velocity $\VEC{v}=(v_1,v_2,v_3)$ is imposed by
an externally applied constant force, and the 
constant symmetric diffusion tensor $\TENSOR{D}$ with components
$D_{ij}$ ($i,j=1,2,3$) characterizes the anisotropic thermal bath
and, possibly, anisotropic particle properties. 
We consider the case of an impenetrable surface
(reflecting boundary) being located at $x_3=0$,
such that $x_3 \geq 0$ measures the height above the surface \cite{note:a}.
The presence of such  hard wall implies a no-flux boundary condition
for the 3-component of the probability flux $v_ip-\frac{\partial}{\partial x_j} D_{ij}p$
at $x_3=0$,
\begin{equation}
\label{eq:noflux}
\left[ v_3 p - \frac{\partial}{\partial x_j} D_{3j} p \right]_{x_3=0} = 0 \, .
\end{equation}

\paragraph*{General results.}
We are looking for a solution of \eqref{eq:FP}, \eqref{eq:noflux}
for a delta-distributed initial density
$p(\VEC{x},0)=\delta(\VEC{x}-\VEC{x}^{(0)})$,
where we require that $x_3^{(0)}>0$.
Without the no-flux boundary condition \eqref{eq:noflux},
the ``free'' (i.e.~unconstrained) solution
$p_{\mathrm{free}}(\VEC{x})=p_{\mathrm{free}}(x_1,x_2,x_3)$
\cite{note:t}
is a trivariate Gaussian with mean
$\mu_i = x_i^{(0)}+v_i t$ and covariance $2D_{ij}t$.
We can  split off the $x_3$ component by rewriting it as
$p_{\mathrm{free}}(x_1,x_2,x_3)=p_{\mathrm{free}}(x_1,x_2|x_3)p_{\mathrm{free}}(x_3)$.
The conditional density $p_{\mathrm{free}}(x_1,x_2|x_3)$
is a two-dimensional Gaussian with mean 
\begin{equation}
\label{eq:mu_tilde}
\tilde{\mu}_i = \mu_i + \frac{D_{3i}}{D_{33}} (x_3 - \mu_3 )
\end{equation}
for the components $i=1,2$, and covariance matrix
proportional to the Schur complement of $D_{33}$ in $\TENSOR{D}$,
\begin{equation*}
\label{eq:cov_cond}
2\tilde{\TENSOR{D}}t =
\frac{2t}{D_{33}}
\left(
\begin{array}{cc}
D_{11}D_{33}-D_{13}^2 		& D_{12}D_{33}-D_{23}D_{13}	\\
D_{12}D_{33}-D_{23}D_{13}	& D_{22}D_{33}-D^2_{23}		\\
\end{array}
\right)
\, .
\end{equation*}
The part
$p_{\mathrm{free}}(x_3)=\frac{1}{\sqrt{4\pi D_{33} t}}\, e^{-\frac{\left(x_3-x_3^{(0)}-v_3 t\right)^2}{4 D_{33} t}}$
represents free diffusion (no boundaries) in one dimension with drift $v_3$
and diffusion coefficient $D_{33}$
(and, accordingly, is normalized over $x_3\in(-\infty,+\infty)$).

In the SM \cite{SM} we show
that the exact time-dependent solution $p(x_1,x_2,x_3)$ of \eqref{eq:FP}
on the half-space $x_3 \geq 0$ with the reflecting boundary condition \eqref{eq:noflux}
retains the (conditional) free diffusion in the $x_1$ and $x_2$ components
$p_{\mathrm{free}}(x_1,x_2|x_3)$,
while the unconstrained $x_3$ component is replaced by
the solution $p(x_3)$ for one-dimensional diffusion on the half-line $x_3 > 0$
with a reflecting boundary at $x_3=0$
\cite{smoluchowski17,chandrasekhar43},
\begin{equation}
\label{eq:p}
p(x_1,x_2,x_3)=p_{\mathrm{free}}(x_1,x_2|x_3)p(x_3)
\, .
\end{equation}
The explicit form of $p(x_3)$ has been derived by Smoluchowski
\cite{smoluchowski17}. It reads
\begin{eqnarray}
p(x_3) & = &
p_{\mathrm{free}}(x_3)
+ \frac{1}{\sqrt{4\pi D_{33} t}} \, e^{-\frac{v_3}{D_{33}}x_3^{(0)}} e^{-\frac{\left(x_3+x_3^{(0)} - v_3 t\right)^2}{4 D_{33} t}}
\nonumber\\
& - & 
\frac{v_3}{2D_{33}} e^{\frac{v_3}{D_{33}}x_3} \, \erfc\left[ \frac{x_3+x_3^{(0)}+v_3 t}{\sqrt{4 D_{33} t}} \right] \, ,
\label{eq:px3}
\end{eqnarray}
where $p_{\mathrm{free}}(x_3)$ is the same expression for free diffusion as before,
but now applied only to the half-line $x_3 \in [0,+\infty)$, and where the additional
terms account for the ``collisions'' of the particle with the wall
\cite{behringer11,behringer12}
($\erfc(x)$ denotes the complementary error function).

\paragraph*{First and second moments.}
Exploiting the factorized form of the solution $p(x_1,x_2,x_3)$ in
\eqref{eq:p}, with
$p_{\mathrm{free}}(x_1,x_2|x_3)$ being a Gaussian, it is possible
to directly compute all the moments of the particle displacement
as a function of time.
Performing the Gaussian integrals over $x_1$ and $x_2$ and using
\eqref{eq:mu_tilde} we obtain for the first moments
\begin{subequations}
\begin{equation}
\label{eq:ave}
\langle x_i \rangle =
\mu_i - \frac{D_{3i}}{D_{33}}\mu_3 + \frac{D_{3i}}{D_{33}} \langle x_3 \rangle
\, ,
\end{equation}
and for the second moments
\begin{equation}
\label{eq:cov}
\langle x_i x_j \rangle - \langle x_i \rangle \langle x_j \rangle =
2 t \tilde{D}_{ij} + \frac{{D}_{3i}{D}_{3j}}{{D}_{33}^2} \left( \langle x_3^2 \rangle - \langle x_3 \rangle^2 \right)
\, ,
\end{equation}
\end{subequations}
with $i,j$ being 1 or 2. 
The remaining integrals over $x_3$
involve combinations of error functions, Gaussians and polynomials
(see \eqref{eq:px3}), but still can be performed analytically.
The resulting, explicit time-dependent expressions for the moments
$\langle x_3 \rangle$ and $\langle x_3^2 \rangle$ are rather lengthy
and are given in the SM \cite{SM}.

Yet, when the drift is pointing towards the wall, $v_3<0$,
the one-dimensional motion in $x_3$ direction reaches a stationary state
for large times given by
\begin{equation}
\label{eq:px3ss}
p_{\mathrm{stat}}(x_3) = \frac{-v_3}{D_{33}} \, e^{\frac{v_3}{D_{33}}x_3}
\, ,
\end{equation}
with stationary mean
$\lim_{t\to\infty}\langle x_3 \rangle = -\frac{D_{33}}{v_3}$
and variance
$\lim_{t\to\infty} ( \langle x_3^2 \rangle - \langle x_3 \rangle^2 ) = \frac{D^2_{33}}{v^2_3}$.
Recalling that $\mu_k= x_k^{(0)}+v_k t$ (for $k=1,2,3$),
the long-time limits of the moments \eqref{eq:ave} and \eqref{eq:cov}
then assume the compact form (with $i,j$ being equal to 1 or 2) 
\begin{subequations}
\label{eq:long}
\begin{eqnarray}
\langle\dot{x}_i\rangle :=
\lim_{t\to\infty} \frac{1}{t} \langle x_i \rangle
& = & v_i-\frac{D_{3i}}{D_{33}}v_3
\, ,
\label{eq:long_drifti}
\\
\lim_{t\to\infty} \frac{1}{2t} \left( \langle x_i x_j \rangle - \langle x_i \rangle \langle x_j \rangle \right)
& = & \tilde{D}_{ij}
= \frac{D_{33}D_{ij}-D_{3i}D_{3j}}{D_{33}}
\, ,
\nonumber\\ &&
\label{eq:long_covij}
\end{eqnarray}
\end{subequations}
and define effective long-term
velocities $\langle \dot{x}_1 \rangle$, $\langle \dot{x}_2 \rangle$ and
effective diffusion coefficients along the surface.

The net average particle velocity \eqref{eq:long_drifti} is the most
striking consequence of our solution:
the long-term average displacements parallel to the surface in 
the directions $x_1$, $x_2$
are not only driven by the drift velocities $v_1$, $v_2$,
but also by the perpendicular component $v_3<0$ in combination
with the elements $D_{31}$, $D_{32}$, respectively, of
the diffusion tensor.
The origin of this motion can be understood as follows.
The drift $v_3<0$ pushes the particle towards the surface
and confines its motion to the close proximity of the plain wall
in an exponential height distribution (see \eqref{eq:px3ss}).
Being forced towards the surface,
the diffusing particle experiences frequent ``collisions'' with this
hard-wall boundary along a preferential direction which is determined by
the anisotropic (but unbiased) thermal environment.
In effect, the anisotropic thermal
fluctuations become rectified,
and induce directed particle motion.
In absence of any drift
components parallel to the surface, $v_1=v_2=0$,
the particle will therefore move over the surface at
a speed and direction that are determined by its diffusion tensor.
This effect can even induce particle migration
against drift forces applied parallel to the surface.

Noise-rectifying particle motion without a systematic force being applied in
the direction of motion is characteristic for ratchet systems
\cite{reimann02,reimann02a,hanggi09},
and is usually generated by broken spatial or time-reversal symmetries in
the applied (potential) forces. Here, no such asymmetric forces are present,
but the overall spatial symmetry is broken by the principal axes of the anisotropic
thermal bath not being aligned with the orientation of the surface.
Indeed, as can be seen from \eqref{eq:long_drifti}, the effect disappears
if the principal axes of the bath are aligned with the surface
($D_{ij}$ diagonal) or
if the bath is isotropic ($D_{ij}$ proportional to the identity).
On the other hand, it might seem from \eqref{eq:long_drifti} that
we can expect noise-rectification to occur even for isotropic
thermal baths if the particle itself is anisotropic
or if there are anisotropies in the viscous properties of the environment
(like, e.g., in the intracellular medium),
because then $\TENSOR{D}$ is generally non-diagonal.
In fact, both cases,
a non-spherical, anisotropic particle as well as anisotropic viscosity,
are characterized by a
friction tensor $\TENSOR{\gamma}$, resulting in a (generally non-diagonal)
diffusion tensor $\TENSOR{D}=\kB T \TENSOR{\gamma}^{-1}$.
Anisotropic friction furthermore
couples the various components of the external constant force
$\VEC{f}=(f_1,f_2,f_3)$ resulting in the drift velocity
$\VEC{v} = \TENSOR{\gamma}^{-1}\VEC{f}$.
The appearance of $\TENSOR{\gamma}^{-1}$ in both,
$\TENSOR{D}$ and $\VEC{v}$, makes the terms proportional to $f_3$
in \eqref{eq:long_drifti} drop out (see \cite{SM}),
such that systematic long-term
drift along the surface can only be induced by the force components
$f_1$, $f_2$ parallel to the surface.
In other words, noise-rectification does not occur if the thermal environment is
isotropic and anisotropic diffusion is only due to anisotropic particle properties
or anisotropic viscosity;
a finding reminiscent of the no-go theorem for ratchet systems with non-constant friction
(see Sec.~6.4.1 in \cite{reimann02}),
and consistent with the fact that an isotropic thermal environment
corresponds to an equilibrium heat bath.

\paragraph*{Experimental proposal.}
In order to illustrate our main result \eqref{eq:long_drifti},
we consider the experimentally realistic situation
\cite{martinez13,mestres14,berut14,berut16}
of a colloidal particle
in an aqueous solution at room temperature $T$,
which is ``heated'' anisotropically
by randomly fluctuating forces applied along
the direction $\VEC{e}_{\sigma}$.
For the sake of generality, we formally keep the tensor properties
of the particle friction $\TENSOR{\gamma}$ when setting up the model
in the following, even though we will exclusively consider
spherical particles in the explicit examples below. 
The particle is pushed towards the plane surface at $x_3=0$ by a constant
external force $\VEC{f}=(f_1,f_2,f_3)$.

Using the model for the anisotropic heat bath,
 put forward and verified in \cite{martinez13},
the equation of (unconstrained) motion for
the Brownian particle can be written as
\begin{equation}
\label{eq:langevin}
\TENSOR{\gamma} \dot{\VEC{x}} =
\VEC{f} + \sigma \VEC{e}_{\sigma} \zeta(t) + \sqrt{2\kB T} \TENSOR{\gamma}^{1/2} \VEC{\xi}(t)
\, ,
\end{equation}
where $\VEC{\xi}(t)=(\xi_1(t),\xi_2(t),\xi_3(t))$ collects three unbiased,
mutually independent Gaussian white noise sources
(with $\langle \xi_i(t)\xi_j(t') \rangle = \delta_{ij}\delta(t-t')$)
which represent the thermal fluctuations, and where $\TENSOR{\gamma}^{1/2}$ is defined via
$\TENSOR{\gamma}^{1/2} \TENSOR{\gamma}^{1/2}=\TENSOR{\gamma}$,
exploiting that $\TENSOR{\gamma}$ is a positive definite tensor.
Finally, $\sigma$ denotes the amplitude of
the anisotropic fluctuations $\zeta(t)$.
Note that in \eqref{eq:langevin} dissipation effects connected to these
fluctuations are assumed to be negligibly small.
It has been demonstrated \cite{martinez13} that $\zeta(t)$ can in very good
approximation be represented as an unbiased delta-correlated white noise,
$\langle \zeta(t)\zeta(t') \rangle = \delta(t-t')$.
In that case, the isotropic thermal noise and the ``synthetic'' directional noise
can be combined into an effective anisotropic thermal environment with different
effective temperatures acting along different directions \cite{martinez13,martinez16,dinis16}
(for details, see the SM \cite{SM}).
Doing so, the equation of motion \eqref{eq:langevin} turns into the equivalent form
\begin{equation}
\label{eq:langevinEff}
\dot{\VEC{x}} = \VEC{v} + \sqrt{2} \TENSOR{D}^{1/2} \, \VEC{\xi}_{\mathrm{eff}}(t) \, ,
\end{equation}
with $\VEC{\xi}_{\mathrm{eff}}(t)$ again being unbiased Gaussian white noise sources,
and with
\begin{subequations}
\label{eq:vD}
\begin{eqnarray}
v_i    & = & (\gamma^{-1})_{ij} f_j
\label{eq:v}
\, , \\
D_{ij} & = & \frac{\sigma^2}{2} (\TENSOR{\gamma}^{-1}\VEC{e}_{\sigma})_i (\TENSOR{\gamma}^{-1}\VEC{e}_{\sigma})_j
 + \kB T \, (\gamma^{-1})_{ij}
\label{eq:D}
\, ,
\end{eqnarray}
\end{subequations}
and $\TENSOR{D}^{1/2}\TENSOR{D}^{1/2}=\TENSOR{D}$.

\begin{figure}[h]
\includegraphics[width=0.9\columnwidth]{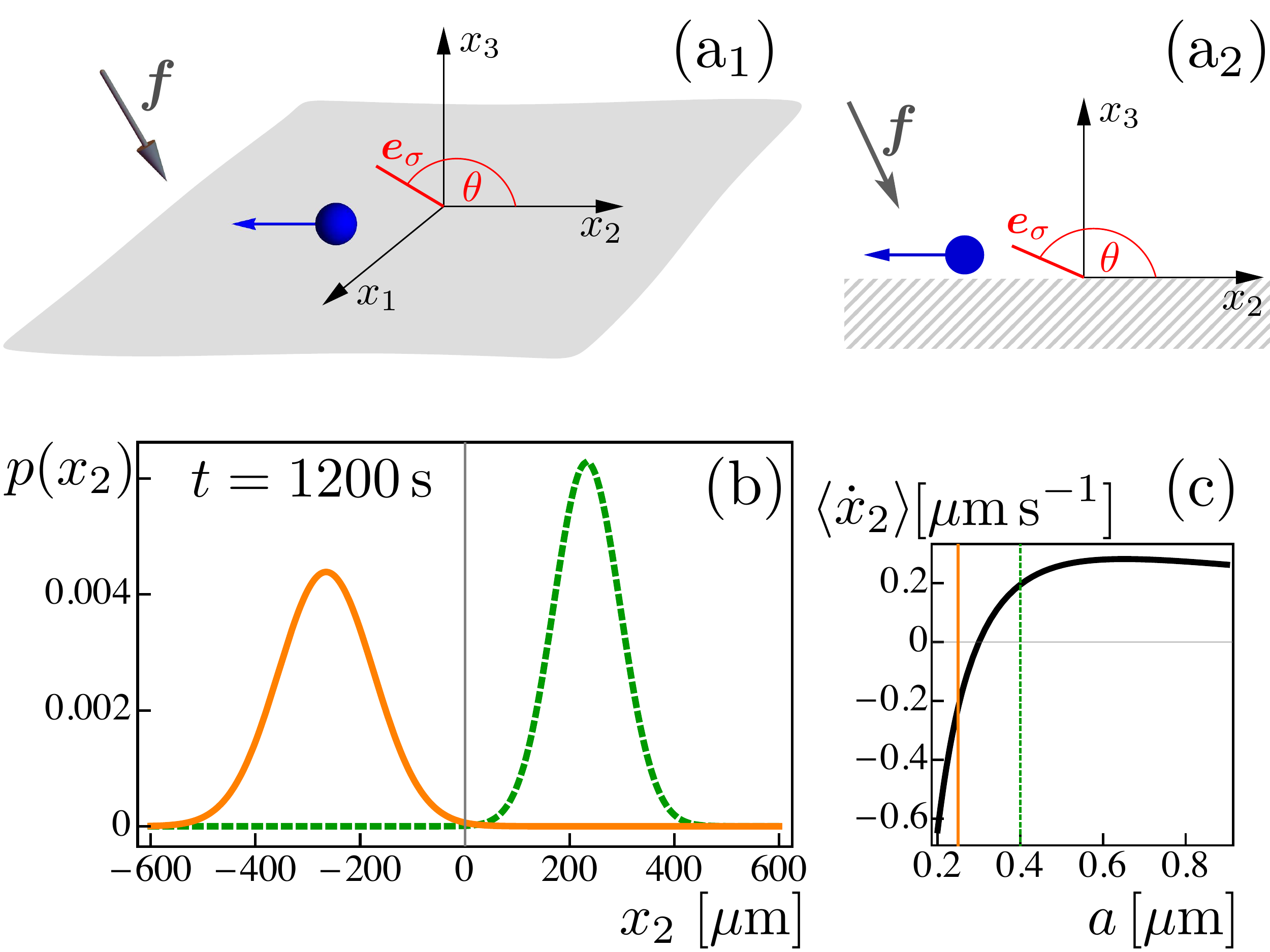}
\caption{%
(a$_1$) and (a$_2$) Illustration of the setup. A Brownian colloid diffuses in proximity to
a two-dimensional surface, the $x_1$-$x_2$ plane at $x_3=0$ (illustrated by the gray areas).
Its average, long-term velocity is indicated by the blue arrow.
The ``synthetic'' noise is applied along the $\VEC{e}_{\sigma}$ direction
in the $x_2$-$x_3$ plane (red bar). The external force $\VEC{f}$ points towards
the surface (gray arrow). In (a$_1$) the full three-dimensional setup is sketched,
(a$_2$) shows the $x_2$-$x_3$ plane in which the ``synthetic''
noise is applied.
(b) Two different spherical particles move along opposite directions on the surface;
solid orange curve: Brownian sphere with radius $\unit[0.25]{\mu m}$;
dashed green curve: Brownian sphere with radius $\unit[0.4]{\mu m}$.
Shown are the probability densities
$p(x_2) = \int_{-\infty}^{+\infty} \mathrm{d}x_1 \int_{0}^{+\infty} \mathrm{d}x_3 \, p(x_1,x_2,x_3)$
for the two particles,
obtained from the exact solution (\ref{eq:p}) at time $t=\unit[1200]{s}$.
(c) Net particle velocity $\langle\dot{x}_2\rangle$, \eqref{eq:drift_explicit2},
as a function of the particle radius $a$. The two particle sizes from (b) are
marked by corresponding solid orange and dashed green lines.
Parameters used in (b) and (c):
$T=\unit[300]{K}$ (room temperature), $T_{\mathrm{kin}}=T+\sigma^2/(2\kB \gamma_0)=\unit[1000]{K}$
(``synthetic'' kinetic temperature \cite{martinez13},
specified for a reference spherical
particle with radius $\unit[0.5]{\mu m}$ and Stokes friction coefficient $\gamma_0$),
$\theta=0.9\pi$, $\VEC{f}=(\unit[0]{fN},\unit[7.5]{fN},\unit[-9]{fN})$.
The friction coefficients of the spherical particles
have been calculated---as an approximation---for an unbounded fluid,
i.~e.~$\tilde{\gamma}=6\pi\nu a$, where $\nu$ is the viscosity of
water at room temperature.
}
\label{fig1}      
\end{figure}

For describing the motion of the Brownian particle
close to the plain surface at $x_3=0$,
the Langevin equation \eqref{eq:langevinEff} (or \eqref{eq:langevin}) does
actually not provide a complete model,
because the hard-core interactions with the reflecting
boundary are not specified, and can in fact not be included
as well-defined interaction forces \cite{behringer11,behringer12}.
We can, however, make use of our
exact solution of the associated Fokker-Planck equation presented above.
In particular, if the deterministic drift is oriented towards
the boundary, the particle's long-term behavior
is determined by \eqref{eq:long} with the velocity field $\VEC{v}$
and the diffusion tensor $\TENSOR{D}$ given in \eqref{eq:vD}.
Specifically,
we consider a spherical particle with a friction tensor
proportional to the identity $\mathbb{I}$,
$\TENSOR{\gamma} = \tilde{\gamma}\mathbb{I}$.
Moreover, we tilt the direction of the ``synthetic''
fluctuations by an angle $\theta$ with respect to the surface,
i.e.~we have $\VEC{e}_{\sigma} = (0,\cos\theta,\sin\theta)$,
by convenient orientation of the $x_1$ and $x_2$ axes such that $\theta$
is the angle between the $x_2$ axis and $\VEC{e}_{\sigma}$
(see Fig.~\ref{fig1}a).
For this set-up, the
explicit expressions for the long-term drift velocities
\eqref{eq:long_drifti} read
\begin{subequations}
\label{eq:drift_explicit}
\begin{eqnarray}
\langle\dot{x}_1\rangle & = & f_1/\tilde{\gamma},
\\
\langle\dot{x}_2\rangle & = &
\frac{1}{\tilde{\gamma}} \left[
f_2 - f_3 \frac{\sin\theta \cos\theta} {\frac{T}{T_{\mathrm{kin}}-T}\frac{\tilde{\gamma}}{\gamma_0} + \sin^2\theta}
\right]
\, ,
\label{eq:drift_explicit2}
\end{eqnarray}
\end{subequations}
where we have introduced the standard definition
$T_{\mathrm{kin}}=T+\frac{\sigma^2}{2\kB \gamma_0}$
for the ``hot'' kinetic (or effective) temperature
\cite{martinez13,martinez16,dinis16}
of a spherical particle with radius $\unit[0.5]{\mu m}$
and Stokes friction coefficient $\gamma_0$ in an unbounded fluid.
The ``hot'' temperature $T_{\mathrm{kin}}$ together with
the direction $\theta$ characterize the anisotropy of the
environment.

From the results \eqref{eq:drift_explicit}
we can make a number of interesting observations
(see also the SM \cite{SM}, where we provide
plots of \eqref{eq:drift_explicit2}):

(i) The average drift
velocity in $x_1$ direction has the trivial form
$f_1/\tilde{\gamma}$,
because the ``synthetic'' noise does not have an $x_1$ component
and thus diffusion in
$x_1$-$x_3$ planes is isotropic
and can not be rectified.

(ii) In an isotropic (equilibrium) thermal bath, when $\sigma = 0$
(implying $T_{\mathrm{kin}}=T$),
net drift along the surface is
only present if there are non-vanishing deterministic force components parallel to the
surface, i.e.~$f_1 \neq 0$ or $f_2 \neq 0$, as already inferred above on more general grounds.

(iii) Likewise, if the ``synthetic'' noise is applied in a direction
parallel or perpendicular
to the surface (i.e.~$\sin\theta=0$ or $\cos\theta=0$), an average drift
velocity over the surface can be induced only by $f_1 \neq 0$ or $f_2 \neq 0$.

(iv) 
The noise rectification effect, which is quantified in \eqref{eq:drift_explicit2}
by the $f_3$-term, depends on the friction coefficient $\tilde{\gamma}$.
It is thus sensitive to particle shape and size,
being stronger for particles with smaller $\tilde{\gamma}$.
For appropriate choices of
the ``synthetic'' noise parameters $\sigma$ and $\theta$,
such that the two terms $f_2$ and
$f_3 \frac{\sin\theta \cos\theta}{\frac{T}{T_{\mathrm{kin}}-T}\frac{\tilde{\gamma}}{\gamma_0} + \sin^2\theta}$
in \eqref{eq:drift_explicit2} have the same sign, 
the rectification effect acts even opposite to the
deterministic force component $f_2$.
Hence, the long-term velocity $\langle\dot{x}_2\rangle$
can change direction when varying particle size
(and thus the friction coefficient $\tilde{\gamma}$),
with the smaller particles moving against the force $f_2$
(see Fig.~\ref{fig1}c).
In other words,
we can always find a combination of $\VEC{f}$ and $\sigma\VEC{e}_{\sigma}$,
which makes two different particle species with different $\tilde{\gamma}$
move into opposite directions on the
surface, such that they become separated with high efficiency
(see Fig.~\ref{fig1}b).
Note that although the ``synthetic'' temperature of the experimental
setup used in \cite{martinez13,martinez16,dinis16} can be made extremely high,
such large temperatures do not necessarily increase the sorting efficiency.
Indeed, for too large $T_{\mathrm{kin}}$, the term
$\frac{T}{T_{\mathrm{kin}}-T}\frac{\tilde{\gamma}}{\gamma_0}$ in
\eqref{eq:drift_explicit2} becomes negligible such that the sensitivity
of $\langle\dot{x}_2\rangle$ for particle properties gets lost.
We furthermore remark that our theory does not take into account
particle-particle interactions and thus describes the dilute limit of
the sorting problem. Moreover, in practice, care has to be
taken by appropriate choices of materials and coatings
that the particles do not stick to the surface.

\paragraph{Concluding remarks.}
The main results of this Letter are the exact, time-dependent solution
$p(t;\VEC{x})$ of the Fokker-Planck equation \eqref{eq:FP} with the
no-flux boundary condition \eqref{eq:noflux} \cite{note:gen},
and its implications for particle diffusion close to a plain surface.
Most notably, we find (see \eqref{eq:long_drifti}) that an anisotropic thermal
environment induces directed particle motion along the boundary even if
no systematic forces are applied in this direction.
To illustrate these results we analyze the average motion of a
Brownian colloid close to a plain surface.
The anisotropic thermal environment
is created by superimposing
externally applied, (almost white) random fluctuations to
the thermal fluid bath, using a technique that has been established
experimentally in recent years
\cite{gomez10,martinez13,mestres14,berut14,dieterich15,martinez15,martinez16_carnot,berut16,martinez16,dinis16}.
In modeling this system with the Fokker-Planck equation
\eqref{eq:FP} and when applying our analytic solution,
we tacitly assume that the friction coefficient of the Brownian particle
is constant, i.e.~independent of particle position.
This idealization does not take into account the
changes of viscous friction with the distance from the surface
due to hydrodynamic interactions \cite{happel83}.
Close to the surface hydrodynamic
friction becomes very large and will slow down the movements of
the particle. Since the particle sorting mechanism we suggest here
occurs in the vicinity of the surface, we therefore expect the sorting
efficiency to decrease
when properly taking into account hydrodynamic effects.
Numerical simulations (see SM \cite{SM}) confirm these expectations,
but also show that our main qualitative finding---systematic
noise-induced transport along the surface in a direction which depends
on particle properties---seems to be robust.
A detailed analysis of hydrodynamic effects will be subject of future work.


\begin{acknowledgments}
\paragraph*{Acknowledgments.}
We thank Hans Behringer for many stimulating discussions,
RE acknowledges financial support
from the Swedish Science Council (Vetenskapsr{\aa}det)
under the grants 621-2012-2982, 621-2013-3956 and 638-2013-9243.
\end{acknowledgments}


\begin{thebibliography}{99}
\bibitem{note:a}
If \eqref{eq:FP} models the motion of a particle of finite size (radius $a$),
the coordinate $x_3$ represents the gap between the surface and the particle, i.e.~the
center of the particle is located at $x_3+a$.
\bibitem{note:t}
Note that we consider the full time-dependent solution $p$
throughout this Letter, but we omit the time-argument
in the following for notational convenience.
\end{thebibliography}
\end{document}